\documentclass[prl, aps, notitlepage, superscriptaddress, showpacs, a4paper,twocolumn,10pt]{revtex4-1} % ,groupedaddress

\usepackage{graphicx}  % needed for figures
\usepackage{bm}        % for math
\usepackage{amssymb}   % for math
\usepackage[latin1]{inputenc}
\usepackage[T1]{fontenc}
\usepackage{lipsum}  
\usepackage[english]{babel}
\usepackage{subfigure}
\usepackage{epsfig}
\usepackage{verbatim}
\usepackage{setspace}
\usepackage{placeins}
\usepackage{amsmath}
\usepackage[normalem]{ulem}
\usepackage{mathrsfs}% bold math
\usepackage{slashed}
\usepackage[usenames, dvipsnames]{color}
\usepackage{graphics}
\usepackage[margin=1in,footskip=0.25in]{geometry} % control over margin
\usepackage[bookmarks=false,pdfborder={0 0 0}]{hyperref}
\usepackage[mediumspace,mediumqspace,squaren]{SIunits}
\usepackage{amsfonts}
\usepackage{hyperref}
\hypersetup{
colorlinks=true,
linkcolor=blue,
citecolor=blue}

\begin{document}

\title{Observation of the Bogoliubov dispersion relation in a fluid of light}

\date{\today}
\author{Q. Fontaine}
\author{T. Bienaim\'e}
\author{S. Pigeon}
\author{E. Giacobino}
\author{A. Bramati}
\affiliation{Laboratoire Kastler Brossel, Sorbonne Universit\'e, CNRS, ENS-PSL Research University, Coll\`ege de France, Paris 75005, France}
\author{Q. Glorieux}
\email[Corresponding author: ]{quentin.glorieux@sorbonne-universite.fr}
\affiliation{Laboratoire Kastler Brossel, Sorbonne Universit\'e, CNRS, ENS-PSL Research University, Coll\`ege de France, Paris 75005, France}

\begin{abstract}

Quantum fluids of light are photonic counterpart to atomic Bose gases and are attracting increasing interest for probing  many-body physics quantum phenomena such as superfluidity. 
Two different configurations are commonly used: the confined geometry where a nonlinear material is fixed inside an optical cavity, and the propagating geometry where the propagation direction plays the role of an effective time for the system. 
The observation of the dispersion relation for elementary excitations in a photon fluid has proved to be a difficult task in both configurations with few experimental realizations.
Here, we propose and implement a general method for measuring the excitations spectrum in a fluid of light, based on a group velocity measurement. We observe a Bogoliubov-like dispersion with a speed of sound scaling as the square root of the fluid density. 
This study demonstrates that a nonlinear system based on an atomic vapor pumped near resonance is a versatile and highly tunable platform to study quantum fluids of light.

\end{abstract}

\maketitle
\paragraph{}
Superfluidity is one of the most striking manifestation of  quantum many-body physics.
Initially observed in liquid Helium~\cite{Kapitza_Helium, Allen_Helium}, the realization of atomic Bose-Einstein condensates (BEC) has allowed detailed investigations of this macroscopic quantum phenomenon exploiting the precise control over the system parameters.
Recently, another kind of quantum fluid made of interacting photons in a nonlinear cavity has brought new perspectives to the study of superfluidity in driven-dissipative systems, with many fascinating developments~\cite{Yamamoto} such as the observation of polariton BEC~\cite{Pfeiffer_Condensation_Polaritons,Kasprzak} and the demonstration of  exciton-polariton superfluidity~\cite{Amo_Superfluidity_Polariton,lerario2017room}.
A different photon fluid configuration, initially proposed by Pomeau and Rica more than twenty years ago~\cite{Pomeau_Rica_Superflow} but long ignored experimentally, relies on the propagation of a intense laser beam through some nonlinear medium.
In this 2D+1 geometry (2 transverse spatial dimensions and 1 propagation dimension analogous to an effective time), the negative third-order Kerr nonlinearity is interpreted as a photon-photon repulsive interaction. Few theoretical works addressing mostly hydrodynamic effects using this geometry have been recently proposed~\cite{Carusotto_Propagating_Geometry, Larre_Defect} and investigated in photorefractive crystals~\cite{Bellec_Defect}, thermo-optic media~\cite{Vocke_Dispersion, Vocke_Geometry} and hot atomic vapors \cite{Kaiser}.

The theoretical framework used to describe quantum fluids of light relies on the analogy with weakly interacting Bose gases where the mean field solution has originally been derived by Bogoliubov~\cite{Bogoliubov_Original_Paper,kohnle2011single}.  
A fundamental property of the Bogoliubov dispersion relation is the linear dependence in the excitation wavevector at long wavelengths (sound-like) and the quadratic dependence at short wavelengths (free-particle like). 
Although this dispersion has been well characterized in atomic BEC experiments~\cite{Jin_Collective_Excitations, Mewes_Collective_Excitations, Onofrio_Superfluid,PhysRevLett.88.120407}, a direct measurement of this dispersion in a fluid of light remains elusive~\cite{Vocke_Dispersion,utsunomiya2008observation}. In this letter, we propose a general method to experimentally access the dispersion of elementary density excitations of a photon fluid. We show that the dynamics of these excitations is governed by a Bogoliubov-like dispersion and that our experimental platform, based on light propagation in hot atomic vapor, is promisings to study hydrodynamics effects emerging in fluid of light systems.
Our experiment settles the question originally asked by R. Chiao two decades ago~\cite{chiao2000bogoliubov}: can one observe sound-like excitations and superfluidity of light  ?

Even if photons in free space are essentially non-interacting particles, engineering an effective photon-photon interactions is possible by exploiting an optical nonlinear process.
In our experiment, the third-order Kerr nonlinearity is induced by the propagation of a near-resonant laser field inside a hot Rubidium atomic vapor.
The sign and the strength of the interactions can be finely tuned by adjusting the laser detuning with respect to the  atomic resonance. 
The atomic density, given directly by the vapor temperature, provides an additional control over the strength of the interactions.
This system has been extensively studied in the context of quantum and nonlinear optics~\cite{glorieux2010double}, but the quantum fluid of light framework gives a better and more complete understanding about the physical phenomena discussed in this letter.
% The propagation of a monochromatic linearly polarized laser field  $E(\textbf{r}_{\perp},z)$ through a nonlinear medium is given, under the paraxial approximation, by the nonlinear Schr\"{o}dinger Equation (NLSE)~\cite{Boyd} :
This framework is derived from the Nonlinear Schr\"{o}dinger Equation (NSE), describing the propagation along the $z$-direction of a monochromatic linearly polarized laser field  $E(\textbf{r}_{\perp},z)$ in a nonlinear medium, when the paraxial approximation is valid:
\begin{equation}
\label{NSE}
i \, \frac{\partial E}{\partial z} = -\frac{1}{2 k_{0}} \nabla_{\perp}^2 E - \left( k_{0} n_{2} \vert E \vert^2 + i \frac{\alpha}{2} \right) E,
\end{equation}
\noindent where $k_{0} = 2 \pi / \lambda_0$ is the laser wavevector ($\lambda_0$ stands for the laser wavelength in vacuum) and $\nabla_{\perp}$ the gradient with respect to the transverse spatial coordinate $\textbf{r}_{\perp} = (x,y)$ .
When the linear absorption coefficient $\alpha$ is negligible and the nonlinear refractive index $\Delta n = n_{2} \, I$ ($I$ represents the laser field intensity) is negative, the NLSE is mathematically analogous to the Gross-Pitaevski equation, describing the dynamics with respect to an effective time $t = z n_{0}/c $ ($c$ stands for the speed of light in vacuum) of a 2D-fluid with repulsive interactions.
Using the Madelung transformation $E(\textbf{r}_{\perp},z) = \sqrt{\rho(r_{\perp},z)} \exp{ \left[ i \, \Phi \left( r_{\perp},z \right) \right] } $, one obtains a coupled system of hydrodynamic equations for the electric field density $\rho$ and phase $\Phi$ :
\begin{equation}
\label{Hydrodynamic_eq_1}
\frac{\partial \rho}{\partial t} + \boldsymbol{\nabla}_{\perp}\cdot \left( \rho \textbf{v} \right) = 0,   
\end{equation}
\begin{equation}
\label{Hydrodynamic_eq_2}
\frac{c}{k_0} \frac{\partial \Phi}{\partial t} + \frac{1}{2} v^2 + c^2 \!  \left( n_{2} \rho - \frac{1}{2 k_{0}^2} \frac{\nabla_{\perp}^2 \sqrt{\rho}}{\sqrt{\rho}}\right)\! = 0,
\end{equation}
where $\textbf{v} = (c/k_{0}) \boldsymbol{\nabla}_{\perp} \Phi$. In this formulation, the laser beam is described as a fluid of density $\rho$ flowing with velocity $\mathbf v$ in the transverse plane.
%%%%%%%%%%%% FIGURE MANIP %%%%%%%%%%%%%%%%%%%%%%%%
%%%%%%%%%%%% FIGURE MANIP %%%%%%%%%%%%%%%%%%%%%%%%
\begin{figure}[t!]\label{Exp_Setup}
\center
\includegraphics[width=\columnwidth]{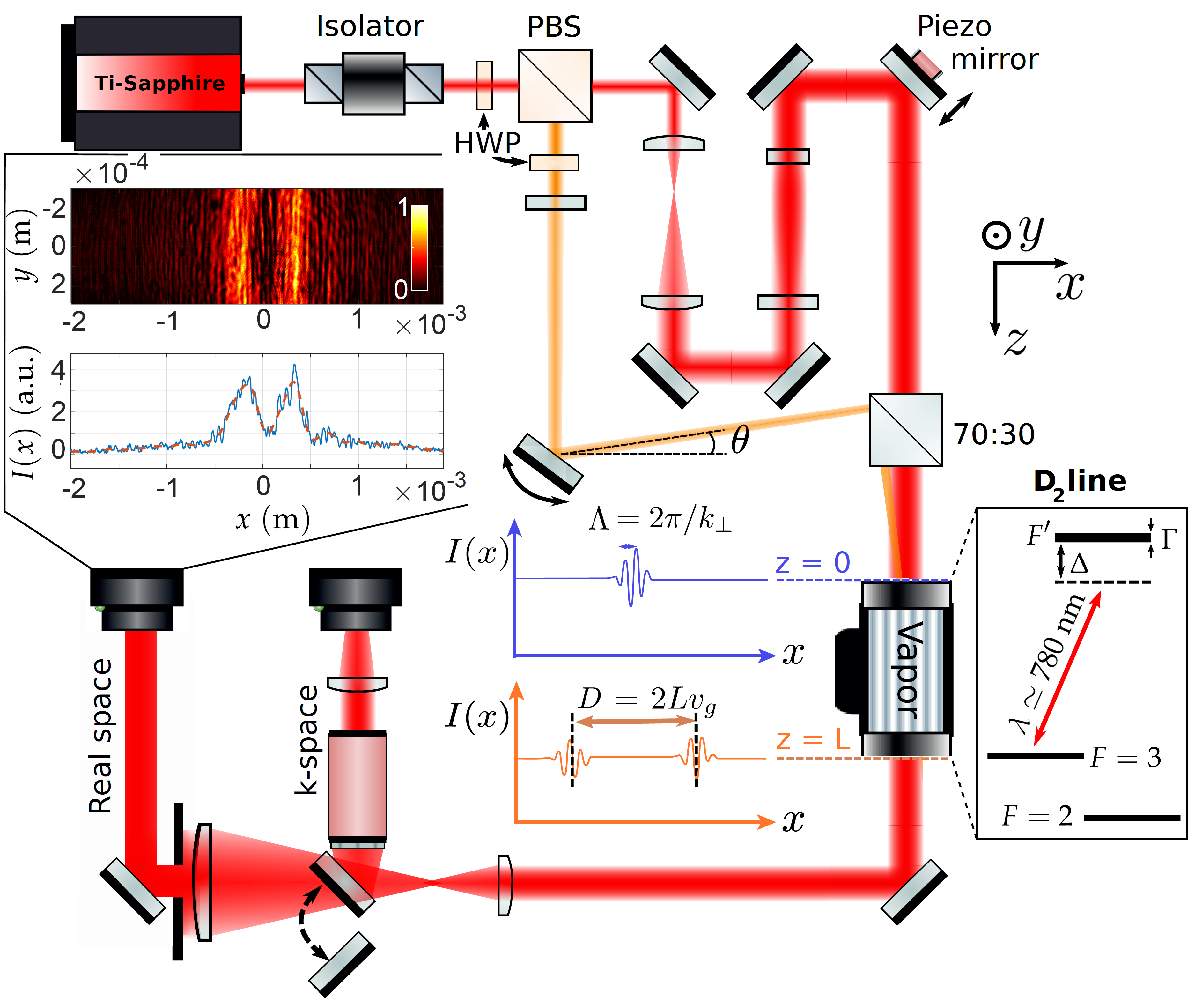} 
\caption{Experimental setup. PBS and HWP stand for Polarized Beam Splitter and Half-Wave Plate respectively. $\theta$ is the angle between the probe (orange beam) and the optical axis defined by the pump (red beam). The probe interferes with the pump and slightly modulates its intensity. Blue inset: integrated intensity profile at the input of the medium ($z=0$). The wavelength $\Lambda$ of the density modulation is given by $2\pi/k_{\perp}$ where $k_{\perp} = k_{0} \sin \theta$. Orange inset: integrated intensity profile at the output of the medium ($z = L$). The distance $D$ between the two wavepackets gives access to the group velocity of the elementary excitations in the transverse plane. The output plane is imaged on a CMOS camera. Inset on the top left: background-subtracted image obtained for $\theta \approx 0$ rad and associated integrated envelope profile (blue: original ; red dotted : high frequency filtered).}
\label{Exp_Setup}
\end{figure}
%%%%%%%%%%%% FIGURE MANIP %%%%%%%%%%%%%%%%%%%%%%%%%
%%%%%%%%%%%% FIGURE MANIP %%%%%%%%%%%%%%%%%%%%%%%%%
The dynamics of the density fluctuations on top of the photon fluid is governed by the Bogoliubov dispersion relation.
For small amplitude modulations moving on a uniform background fluid at rest, the set of hydrodynamic equations can be linearized assuming $\rho = \rho_{0} (z)+ \delta \rho (\textbf{r}_{\perp},z)$ and $\textbf{v} = \delta \textbf{v} (\textbf{r}_{\perp},z)$. By taking the transverse gradient of Eq.~\eqref{Hydrodynamic_eq_2}, keeping the first order terms in the expansion and using Eq.~\eqref{Hydrodynamic_eq_1}, one can derive an equation for $\delta \rho$ only.
For a plane-wave density fluctuation mode $\delta \rho$ of wave vector $\textbf{k}_{\perp}$, the associated response frequency $\Omega_{B}$ will follow the dispersion relation below :  
\begin{equation}
\label{Dispersion_Relation}
\Omega_B(\textbf{k}_{\perp}) = c \sqrt{\vert \Delta n \vert \, \textbf{k}_{\perp}^2 + \left( \frac{\textbf{k}_{\perp}^2}{2 k_{0}} \right)^2}.
\end{equation}
When the wavelength $\Lambda=2\pi/\vert\textbf{k}_{\perp}\vert$ of the modulation is longer than the \textit{healing length} $\xi = \frac{\lambda}{2} \sqrt{ \frac{1}{\vert \Delta n \vert}}$, the dispersion relation becomes linear and the modulations propagate as sound waves.
This regime is characterized by the sound velocity $c_s = c \sqrt{\vert \Delta n \vert}$, which only depends on the nonlinear index of refraction $\Delta n$.
Conversely, when $\Lambda \gg \xi$, the dispersion relation becomes quadratic which is similar to the free propagating particle one.
The existence of a sound-like regime in the Bogoliubov dispersion is, according to the Landau criterion, a necessary condition for observing superfluid flow of light, as discussed in~\cite{Chio_Superflow}.

Observing the sound like-regime of the Bogoliubov dispersion relation has been proposed in~\cite{Carusotto_Propagating_Geometry} and first attempted in~\cite{Vocke_Dispersion} for propagating geometries. 
The approach used in~\cite{Vocke_Dispersion} relies on the measurement of the phase velocity difference between plane wave density modulations propagating at a given transverse wavevector $k_\perp = 2 \pi / \Lambda$ on top of a high and a low density photon fluid.
The photon fluid is obtained by sending a wide laser beam through a self-defocusing nonlinear medium; the fluid density is then given by the light intensity. 
The small amplitude plane wave density modulation is produced by interfering this first beam with a wide and weak probe field, propagating with a small angle with respect to the optical axis.
In this configuration, however, a \textit{conjugate} wave propagating in the opposite transverse direction ($-\textbf{k}_{\perp})$ is spontaneously generated at the linear/nonlinear interface~\cite{Larre_Quench} .
Probe and conjugate overlap and interfere, which strongly alters the phase shift measurement used to determine the dispersion relation. Moreover, the large nonlinearity needed to observe the sonic dispersion makes extracting the dispersion relation from this measurement rely on a complex numerical inversion \cite{Larre_Phase}. 
On the contrary, we present a direct and intuitive method to extract the dispersion relation for arbitrary modulation wavelengths.
%%%%%%%%%%%% FIGURE SIMU %%%%%%%%%%%%%%%%%%%%%%%%%
%%%%%%%%%%%% FIGURE SIMU %%%%%%%%%%%%%%%%%%%%%%%%%
\begin{figure}[h]
\center
\includegraphics[height=8.5cm]{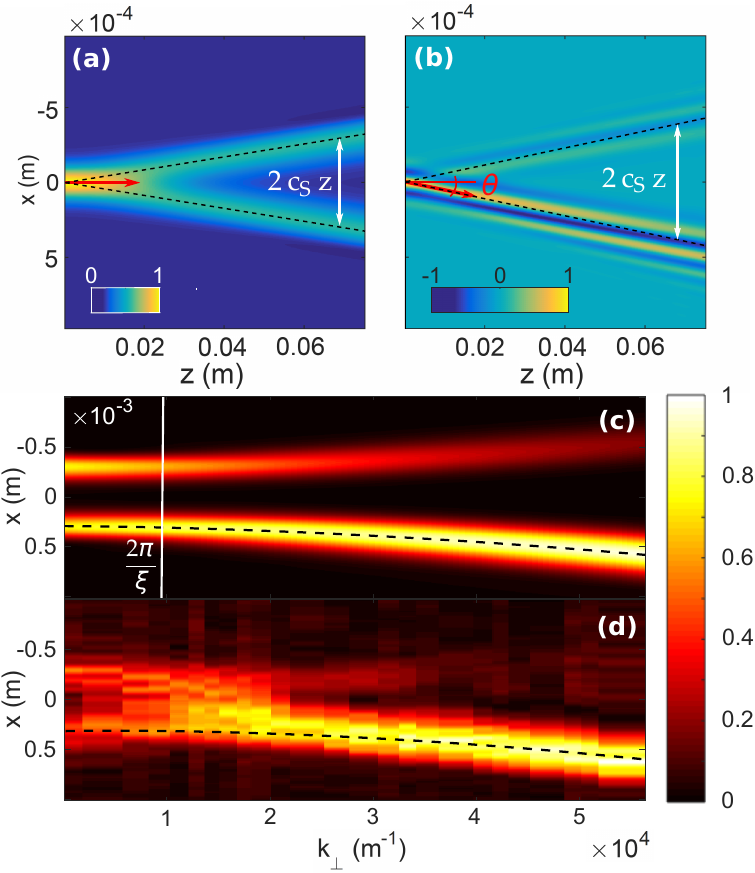}
\caption{Simulation of the propagation of a weak transverse modulation through a nonlinear medium (a) with zero transverse speed ($\theta = 0$ rad).
The modulation generates two counter propagating modes (Bogoliubov modes) at the medium interface and get amplified until they separate from each other. 
The wavepacket is not spreading along propagation due to the non-dispersive regime (sound-like behavior).
(b) Same as (a), for an incident probe at $\theta \! = \! 5\times \! 10^{-3}$ rad (high transverse speed). 
Interference fringes appear and the wavepacket spreads.
(c) simulation of the intensity profile envelope in the output plane for different probe wavevector. 
Dashed black: group velocity given by Eq.~\eqref{Dispersion_Relation}.
(d) experimental data.
The parameters in (a), (b) and (c) are those used to obtain the experimental data (d) : $\lambda_{0} = 780$ nm, $\Delta n = 1.3 10^{-5}$ and $\omega^{p}_{x} = 180$ \micro m. The absorption coefficient $\alpha$ was set to 0 in numerical simulations. Colorbars encode the light intensity in arbitrary units.}  
\label{fig:1D_simulation}
\end{figure}
%%%%%%%%%%%% FIGURE SIMU %%%%%%%%%%%%%%%%%%%%%%%%%
%%%%%%%%%%%% FIGURE SIMU %%%%%%%%%%%%%%%%%%%%%%%%%
Our approach is based on the measurement of the group velocity of a small amplitude Gaussian wavepacket travelling on top of the photon fluid with the transverse wave vector $k_\perp$. 
This wavepacket is designed by interfering the wide and intense beam forming the fluid (at $\textbf{k}_{\perp} = \textbf{0}$) with a Gaussian probe at $\textbf{k}_{\perp} = k_{0} \sin{\theta}\, \textbf{e}_x$, as depicted in Fig.~\ref{Exp_Setup}.
At the entrance of the nonlinear medium, the effective photon-photon interaction constant undergoes a sudden jump along the optical axis. 
Two counter-propagating wavepackets are spontaneously created from the initial Gaussian perturbation and evolve over the effective time $t$ through the nonlinear medium, with a transverse group velocity $\pm v_{g}$. 
The separation between these two modulations at a given propagation distance $z$ (\emph{i.e.} at given time $t)$, is a direct measurement of the group velocity.
In the output plane $(z \! = \! L)$, this distance is given by $D(k_{\perp}) = 2 L v_{g}(k_{\perp}) $. 
The dispersion relation $\Omega_{B}(k_\perp)$ is reconstructed by scanning the wavevector of the modulation $k_\perp$ (tuning the angle $\theta$ between pump and probe) and integrating the group velocity $v_{g}$: $\Omega_{B}(k_\perp) = \int_0^{k_\perp} v_g(q) \,$dq.

%In order to illustrate our method, we solve numerically the nonlinear evolution of the transverse electric field (pump + probe) with the second-order split step Fourier method, for one transverse spatial dimension only (1D+1 geometry), taking advantage of symmetries in the flat fluid density situation (infinitely wide background beam).
In order to illustrate our method, we solve numerically the NSE Eq.~\eqref{NSE} to get the evolution of the transverse electric field (pump + probe). We use the second-order split step Fourier method, for one transverse spatial dimension only (1D+1 geometry) to take advantage of symmetries in the flat fluid density situation (infinitely wide background beam).
The probe waist is located in the entrance plane at $z=0$; its width $\omega^{p}_{x}$ is the same as the one used in the experiment.
For all the density plots in Fig. \ref{fig:1D_simulation}, the uniform background intensity has been subtracted. 
The evolution of the two counter-propagative modulations generated at the entrance of the nonlinear medium is shown in Fig.~\ref{fig:1D_simulation}(a) for zero initial transverse speed and presents a sound like-behavior (no spreading of the wavepacket).
The Fig.~\ref{fig:1D_simulation}(b) is obtained for on high transverse initial speed modulation which behaves like a free-particle. 
Notice that for small incident angle, corresponding to zero initial transverse speed, the two modulations generated at the entrance of the nonlinear medium acquire a non-zero opposite transverse speed. 
This nonlinear refraction law comes from the linear nature of the dispersion for $k_{\perp} \ll 2\pi/\xi$ \cite{Larre_Quench}, which is counter-intuitive from the linear optics perspective. 
The envelope of the intensity profile in the output plane is presented as a function of the probe wavevector in Fig. \ref{fig:1D_simulation}(c), on top of the experimental results in Fig. \ref{fig:1D_simulation}(d). 
The black dotted line represents the theoretical group velocity $v_{g}$, obtained by taking the derivative of Eq.~\eqref{Dispersion_Relation}. 
The distance between the two wavepackets is constant for $k_{\perp} \lesssim 2\pi/\xi$ (linear dispersion; constant $v_g$) and linearly increase for larger $k_{\perp}$ (quadratic dispersion; $v_g\propto k_{\perp}$).

%%%%%%%%%%%% FIGURE VG vs K %%%%%%%%%%%%%%%%%%%%%%%%%
%%%%%%%%%%%% FIGURE VG vs K %%%%%%%%%%%%%%%%%%%%%%%%%
\begin{figure}[]
\center
\includegraphics[height=8.5cm]{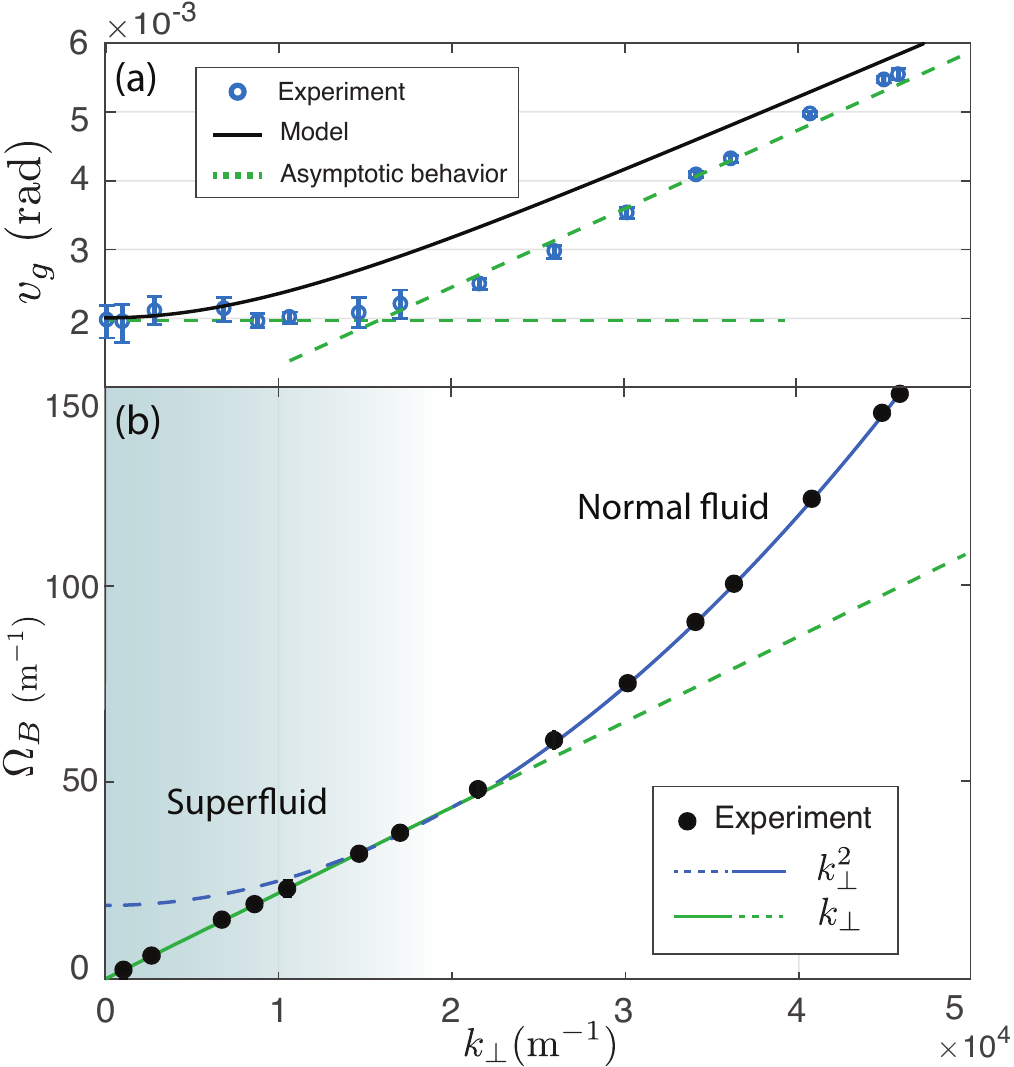}
\caption{a) Group velocity as a function of the transverse wavevector $k_\perp$. The circles represent the experimental data obtained for $P=175$~mW. 
The theoretical model is plotted in solid dark (parameters extracted from an independent measurement of the nonlinearity -- see text for details). 
The dashed green lines are the asymptotic behaviors: constant group velocity at small $k_{\perp}$ and linear increase at large $k_{\perp}$.
b) Dispersion relation obtained after integration of the group velocity. 
Linear (green) and parabolic (blue) dispersion curves are plotted as a reference. 
$\Omega_B$ has the dimension of an inverse length}
\label{Dispersion}
\end{figure}
%%%%%%%%%%%% FIGURE VG vs K %%%%%%%%%%%%%%%%%%%%%%%%%
%%%%%%%%%%%% FIGURE VG vs K %%%%%%%%%%%%%%%%%%%%%%%%%

The experimental setup is shown in Fig.~\ref{Exp_Setup}. 
A continuous-wave Ti:Sapphire laser beam is split into two beams: a low power probe and a high power pump. 
The pump is expanded twice before being focused in the center of the nonlinear medium with two cylindrical lenses to create an elliptical beam with a width along $x$ of $\omega^{0}_{x} \approx 3.2$~mm and a width along $y$ of $\omega^{0}_{y} \approx 300$~\micro m. 
The pump intensity in the central region can thus be considered as spatially uniform along $x$. 
The Rayleigh length $z_{R,y}^{0}$ associated to $\omega^{0}_{y}$ is $37$~cm, which is five times longer than the length of the nonlinear medium. 
Therfore, We can safely consider the pump beam as being collimated and neglect its divergence along the propagation direction.
The probe is directly focused with a cylindrical lens on the entrance of the nonlinear medium in order to get a flat phase profile.
This beam is elliptically elongated along the $y$ direction.
We set the major axis width $\omega^{p}_{y}$ to 1700 ~\micro m and $\omega^{p}_{x}$ to $180 \pm 10$ ~\micro m in order to properly separate the Gaussian wavepackets in the output plane and conserve the probe collimation along its propagation in the nonlinear medium ($z_{R,x}^{p} \! \approx \! 13$ cm).
% We ensure that the probe intensity at the beam waist stays a hundred times smaller than the pump one.
We fix the probe intensity at its waist to $1\%$ of the pump intensity.
This pump/probe cross configuration enables us to both get closer to the 1D case and to increase the integration range along $y$.  
The angle $\theta$ between pump and probe in the $(xz)$ plane can be finely tuned thanks to a piezo-actuated mirror mount.

\indent Both beams propagate through a $L = 7.5$~cm long cell, filled with an isotopically pure $^{85}$Rb vapor.
The cell is heated up to 150\degree C by an oven designed to reduce air turbulence close to the cell windows.
Adjusting the temperature allows us to control the atomic density and therefore the strength of the optical nonlinearity.
In our case, this optical nonlinearity is obtained by tuning the laser frequency close to the $^{85}$Rb $D_{2}$ resonance line, composed of 2 hyperfine ground states ($F = 2,3$) and 4 hyperfine excited states ($F' = 1$ to 4). Since the laser is highly red-detuned from the $F = 3 \rightarrow F'$ transitions ($\Delta=-6$~GHz), the Doppler broadening can be safely neglected and the negative nonlinear susceptibility is close to the one of an effective two-levels system with only one excited state of decay rate $\Gamma = 6.06$~MHz.   
At these temperature and detuning, the transmission coefficient of the laser beams through the cell is above 70$\%$, which allows one to neglect multiple scattering of light (atom-light interaction processes are mainly dominated by Rayleigh scattering events).
In comparison to~\cite{Vocke_Dispersion}, we can consider that the nonlinear interactions are local, as long as the length scale of the ballistic transport of excited atoms stays much shorter than the healing length, which is the case at that temperature.  

\indent The output plane of the cell is imaged on a CMOS camera.
A microscope objective can be flipped on the beam path to image the far-field (\emph{i.e.} k-space) and measure the probe transverse wavevector $k_{\perp} = k_{0} \sin \theta$.
For every angle $\theta$, the pump intensity (background), the probe intensity and the $k$-space are captured.  
The relative phase between pump and probe is scanned over $2 \pi$.
$40$ background-subtracted images are taken during the phase scan. They are then integrated over one hundred pixels around $(Ox)$ and averaged in absolute values. 
Averaged images before integration are shown in inset of Fig.~\ref{Exp_Setup}(b).  
The distance $D$ between the counter-propagating wavepackets is estimated by performing a two-Gaussian fit for small $k_{\perp}$ \emph{i.e.} when the conjugate beam is visible. For large $k_{\perp}$, the conjugate is not sufficiently amplified anymore and $D$ is directly measure from the distance between the input and output positions of the probe beam.
In order to fully characterize our system, the third order Kerr susceptibility $n_{2}$ is calibrated independently by measuring the self phase accumulated by a slightly defocusing Gaussian beam propagating through the cell~\cite{Rings_Focusing_Defocusing_Exp, Rings_Rb}. 
With the detuning and temperature reported earlier, we found $n_{2} = 3.1 \pm 0.2 \, 10^{-11}$ m$^{2}/$W.  

%%%%%%%%%%%% FIGURE cs vs I %%%%%%%%%%%%%%%%%%%%%%%%%
%%%%%%%%%%%% FIGURE cs vs I %%%%%%%%%%%%%%%%%%%%%%%%%
\begin{figure}[]
\center
\includegraphics[width=\columnwidth]{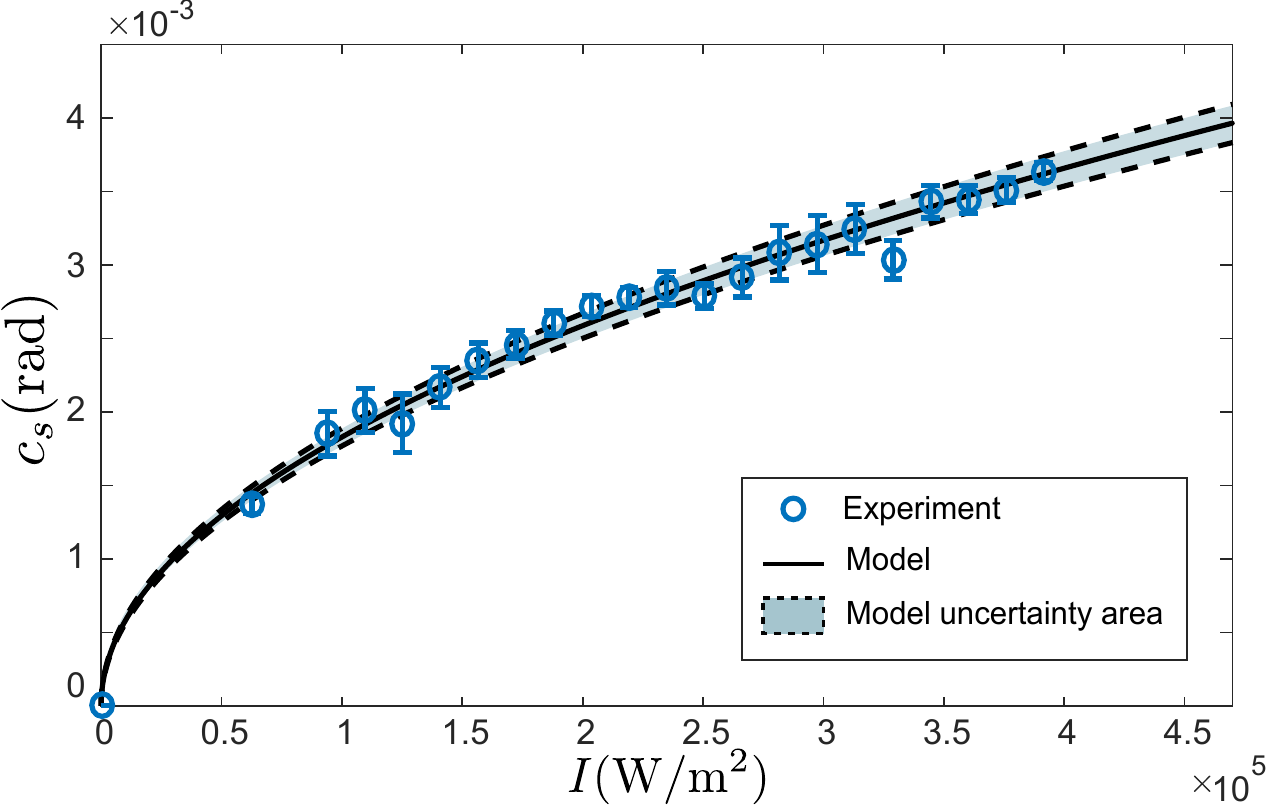} 
\caption{Speed of sound $c_s$ as a function of the pump intensity. Due to the 2D+1 geometry, the speed of sound has the dimension of an angle.
Data is plotted in blue dots.
The light intensity corresponds to the fluid density, therefore a scaling as square-root is expected as plotted in black solid. 
No free parameters are needed as the nonlinearity has been measured independently. 
Uncertainty area in light blue is extracted from this independent measurement. }
\label{csvsI}
\end{figure}

%%%%%%%%%%%% FIGURE cs vs I %%%%%%%%%%%%%%%%%%%%%%%%%
%%%%%%%%%%%% FIGURE cs vs I %%%%%%%%%%%%%%%%%%%%%%%%%

\indent The experimental group velocity and dispersion relation as a function of the probe transverse wavevector are shown in Fig.~\ref{Dispersion}. 
The pump power was set at $175$~mW leading to a nonlinear refractive index $\Delta n$ of $3.9 \, 10^{-6}$.
Two different regimes can be identified on Fig.~\ref{Dispersion}(a): a constant group velocity at low $k_{\perp}$ and a linear increase at larger $k_{\perp}$.
The group velocity clearly goes toward a non-zero value when $k_{\perp} \rightarrow 0$, breaking the linear trend characteristic of the standard free-particle dispersion. 
The theoretical model plotted in Fig.~\ref{Dispersion} is obtained with no free fitting parameters. 
The offset at large $k_{\perp}$ between the model and the experimental data results from constructive interferences between the two non-fully separated wavepackets, as can be seen on the experimental data of Fig.~\ref{fig:1D_simulation}(d) around $k_{\perp} \! \sim \! 1.5 \, 10^4$~m$^{-1}$ (the envelope intensity significantly increases in between them leading to a systematic under-estimation of the distance $D$ by the two-Gaussians fit). After propagation in the cell, the counter-propagating wavepackets have respectively accumulated the phase $\pm \Omega_{B}(k_{\perp}) \, L$. Constructive interferences occur when $\Omega_{B}(k_{\perp}) \, L = n \pi $ ($n$ is a positive integer) \emph{i.e.} for $k_{\perp} \! \sim \! 1.8 \, 10^{4}$ m$^{-1}$ when $n \!= \! 1$. 
%This value fits the position of the crossing point between the two asymptotic regimes. 
This value gives the position of the end of the plateau-like regime at low $k_{\perp}$.
Another constructive interference should occurs for $k_{\perp} \! = \! 0$ (when $n \!= \! 0$), but as both envelopes have the same amplitude in that case, the two peaks are still disentangle (see inset of Fig.~\ref{Exp_Setup}).

%More importantly, our data follow the two asymptotic behaviours of the Bogoliubov dispersion plotted in dashed green in Fig~\ref{Dispersion}(b): a constant group velocity at low $k_{\perp}$ and a linear increase at larger $k_{\perp}$.
More importantly, the dispersion relation of Fig.~\ref{Dispersion}(b) guarantees that light is superfluid in our experiment. 
Indeed, the Landau criterion for superfluidity defines a critical transverse speed $v_c=\underset{k_{\perp}}{\text{min}}\left[\frac{\Omega_B}{k_{\perp}}\right]$ for the photon fluid, below which the emission of sound-like excitations is not possible anymore. 
In our case, $v_c = c_s > 0$ and one could observe superfluid flow of light around a defect if its transverse velocity $v$ (measured in the defect frame) was lower than $c_s$.
To investigate the sonic regime, we set the probe wavevector to zero and record directly the sound velocity as function of the background fluid density (the pump intensity $I$).
The experimental data are shown in Fig.~\ref{csvsI} (blue circles). We observe that the speed of sound scales with the square-root of the fluid density (plotted in black solid) as expected. 
It is worth mentioning that, once again, the third order Kerr susceptibility measured independently, sets the only parameter of the theoretical model.

%\paragraph{Conclusion}
In conclusion we have reported two important experimental results: first we measured the dispersion relation for small amplitude density fluctuations, which shows a linear trend at low wavevector, characteristic of a superfluid. 
We have then assessed the associated sound velocity for different fluid of light densities and obtained a scaling law analogous to the hydrodynamic prediction. 
This settles the question initially asked by Chiao about the possibility to observe a superfluid dispersion in a photon fluid.
These results open a wide range of possible experiments in hydrodynamics with light using a novel versatile platform based on hot atomic vapors.

\paragraph{Acknowledgments.}
The authors want to thank Daniele Faccio for stimulating discussions at the early stage of the project, Iacopo Carustto and Pierre-Elie Larr\'e for important remarks on the physics of Bose gases and superfluidity.

\bibliography{mybib}{}
\bibliographystyle{apsrev4-1}

\end{document}